# Tunable hinge skin states in a hybrid skin-topological sonic crystal

Yu-Xin Fang [1,2], Wen-Hao Zhu[1,2], Yu-Ping Lai[1,2], Yongyao Li[1,2] and Shi-Qiao Wu[1,2†]

[1]*School of Physics and Optoelectronic Engineering, Foshan University, Foshan 528000, China*
[2]*Guangdong-Hong Kong-Macao Joint Laboratory for Intelligent Micro-Nano Optoelectronic Technology, Foshan University, Foshan 528000, China*

[†]Correspondence should be addressed to: sqwu@fosu.edu.cn

Higher-order topological states in sound have played a pivotal role in understanding the intricate physics underlying sound transport, giving rise to new strategy of manipulating sound. Here we report tunable structure for hinge skin states in a non-Hermitian acoustic metamaterial with hybrid skin-topological effect. Our finding shows that when on-site gain and loss are exquisitely introduced into acoustic topological insulators, chiral edge modes in Hermitian counterpart would respectively become amplified or attenuated at zigzag boundaries. If adjacent gain and loss boundaries are intentionally constructed, hinge skin states would take place at their intersections. By strategically combining non-Hermitian and topological physics, we successfully reveal how higher-order hinge modes originate from lower-order surface states and demonstrate flexible acoustic steering in tunable non-Hermitian blocks. Our findings unveil that skin-topological effect may hold significant applications in designing interesting acoustic devices with unconventional functions such as multidimensional acoustic control and accurate energy harvesting.

## I. INTRODUCTION

Sonic crystal (SC) is an artificially periodic structure designed to engineer sound wave. In previous works reporting SCs, efforts have been devoted to the study of acoustic imaging [1], cloaking [2, 3] and wavefront shaping [2]. With the flourishing development of novel topological physics originally predicted in electronic materials, acoustic topological states subsequently have also been extensively explored and successfully observed [4-6]. Sonic topological insulators offer an avenue for disorder-immune sound transport, which hold promise for practical application. For instance, acoustic Chern insulators with synthetic gauge flux have been implemented in acoustic networks consisting of connected ring cavities by introducing non-homogenous flow in the background [4-6]. As a result, quantum Hall effect can be intimately mimicked. Besides, quantum spin Hall effect can also be emulated by folding energy band through breaking original crystalline symmetry [7-9]. Valley topological states, as another widely explored phases, arise in acoustic crystals by perturbing inversion symmetry[10-12]. Among all these acoustic topological materials, robust routing effect of topologically protected edge state is one of the most remarkable hallmarks. And the gapless edge states at sample boundaries can be characterized by distinct topological invariants. Over the past several years, higher-order topological insulators (HOTIs) [13-19] and higher-order topological semimetals (HOTSs) [20-23] as new types of topological materials have attracted considerable interest. Distinct from conventional topological candidates, higher-order topological materials support edge states with two or more dimensions lower than those of their bulks, which have been versatile platforms to engineer sound.

The aforementioned review of acoustic topological phases is limited to Hermitian structures, which means the systems have no energy exchange with their ambient environment. However, in the natural world, dissipation is ubiquitous and non-Hermitian systems are common. Therefore, it is very reasonable to consider non-Hermitian sonic metamaterial from practical implication. Overall, there are two approaches to achieve non-Hermitian topological phases. One way is utilizing the nonreciprocal hopping between neighboring sites in sample with finite length, in which considerable eigenstates would accumulate at the ending of the boundaries and form the so-called non-Hermitian skin effects (NHSEs) [24-29]. However, this method would suffer fundamental challenge when performing in various experimental proposals such as optical lattices [30], acoustic lattices [31] and electrical circuit [32]. For the other hand, gain and loss as imaginary on-site potential can be also adopted to attain novel non-Hermitian topological phases [33-35]. In contrast to nonreciprocity, complex potential is more feasible in realistic experiments. Up to now gain and loss has produced a great deal of new physics including passive *PT*-symmetry breaking [36, 37], pinning of zero energy [38, 39] and hatch of topological insulating phases [33, 34, 40].

Recently, the boundary physics in nonequilibrium open systems are further enriched by higher-order skin effects (HOSEs) [31, 41, 42]. For the first-order skin effects, a system with size $L \times L$ hosts the order of $O(L \times L)$ skin modes at the boundaries. In contrast, for the second-order skin effects, the scale of $O(L)$ skin modes would appear at the corners and most modes remain bulk states, which are critically dictated by the intrinsic topology of the non-Hermitian system. More recently, a very limited number of corner states accompanying with a great deal of bulk states can also arise in hybrid skin-topological systems [43-46]. In such case, skin effect and bulk topology are concurrent. The bulk topology determines topological edge states. And the corner skin states can be viewed as skin modes of the edge states as long as nonzero gauge fluxes at the boundaries persist. Thus, hybrid skin-topological effects demand higher requirement. So far, the exotic phenomena that emerge from hybrid skin-topological effect have been observed in a variety of systems including tight-binding lattice models [43-45, 47] and continuous photonic Chern insulators [46], as well as Floquet systems [48, 49]. Recently, we noticed that two works [50, 51] have reported hybrid skin-topological effects of air sound in experiment. But they mainly focus on the property of point gap topology from the perspective of non-Hermitian in Kane-Mele model. Theoretical models have been utilized to elaborate the forming mechanism of hybrid skin-topological effects, while their relation with the lower order states is still elusive. Here, we exploit non-Hermitian SC in layer-stacked Haldane model to demonstrate the anomalous surface states and their impact on higher-order skin-topological effects. The focus is the intriguing feature of gain and loss edge modes and how they induce higher order skin modes at the corner by assembling gain and loss boundaries. Secondly, we also highlight the adaptability of hybrid skin-topological effect in lattice arrays with a variety of configurations. These are the outstanding points in our work distinct from the cited references [50,51].

In this paper, we subtly design and carefully demonstrate a three-dimensional (3D) non-Hermitian acoustic metamaterial that host gain and loss topological modes induced by the joint interaction of non-Hermiticity and synthetic gauge flux. Without the non-Hermiticity, the acoustic metamaterial is a Weyl crystal, whose 3D band structure contains topologically nontrivial degeneracies named Weyl points [52-55]. The chiral coupling between stacked layers enables inversion symmetry broken, thereby leading to topological Weyl physics for the emergence of boundary-bound modes. For each fixed momentum component $k_z$, the 3D Weyl crystal can be viewed as a two-dimensional (2D) acoustic Chern insulator and its boundary-bound mode present as a chiral surface state with topological robustness against various perturbation. With the introduction of gain and loss into the acoustic resonators, the non-Hermitian topological physics are triggered by Weyl rings rather than Weyl points. In such case, chiral surface modes become amplified or attenuated at zigzag boundaries, which are respectively referred to as gain and loss surface states. The non-Hermitian-induced gain and loss modes carry nonzero orbital angular momentum (OAM) tied to their transport directions. Furthermore, the decay rate of two non-Hermitian surface bands varies strikingly. In contrast, chiral transport of edge dispersions still sustains in armchair-shaped supercell. Consider hexagonal and quadrangular samples with zigzag boundaries, we found that corner states occur at the intersected joints of gain and loss surfaces. If gain and loss are interchanged, skin modes will accumulate towards the other three hinges. This hybrid skin effect is directly associated to the chiral surface current, which is induced by complex on-site potential at zigzag boundaries. Hence, the hybrid skin-topological modes arise from the intricate interplay of gauge flux and non-Hermiticity at zigzag boundaries. And the hinge skin states are the byproducts of hybrid skin-topological effect. To demonstrate this, we further design hexagonal and quadrangular samples with empty holes within the central bulk to observe the hinge skin modes, demonstrating the hybrid skin-topological effect. Triangular sample is also exemplified to illustrate the importance of boundaries for the emergence of hinge skin modes. At last, defects are deliberately introduced by removing the complex on-site potentials at the boundaries to demonstrate the topological robustness of hinge skin states against local disorders.

This work is organized as follows. In Sec. II, we first design a 3D layer-stacked model and demonstrate its equivalence with the 2D Haldane model. Then we implement the 3D theoretical model in realistic SC structure and study its band structures. In Sec. III, we investigate the gain and loss surface dispersions of the 3D SC under different boundary conditions and analyze their transport on the surfaces. In Sec. IV, we construct tunable acoustic arrays and reveal the underlying mechanism of hybrid skin-topological effect. Also, we demonstrate the robustness of hybrid skin-topological effect against local defects. In Sec. V, we summarize this work and anticipate the potential application and effect of our work.

## II. LAYER-STACKED SONIC HALDANE MODEL AND ITS BAND STRUCTURE

We first construct layer-stacked Haldane model with chiral coupling between distinct layers and staggered on-site potential as the tight-binding model showcases in Fig. 1(a). It can be well characterized by the kernel Hamiltonian as follows:

$$H = \begin{pmatrix} i\gamma + t_2\beta_2(k_x, k_y, k_z) & t_1\beta_1(k_x, k_y) \\ t_1\beta_1(-k_x, k_y) & -i\gamma + t_2\beta_2(k_x, k_y, -k_z) \end{pmatrix} \quad (1)$$

wherein $\beta_1(k_x, k_y) = e^{ik_x/\sqrt{3}} + 2\cos(k_y/2)e^{-ik_x/2\sqrt{3}}$ and $\beta_2(k_x, k_y, k_z) = 2\cos(k_y - k_z) + 4\cos(\sqrt{3}k_x/2)\cos(k_z + k_y/2)$. The hopping between nearest neighboring (NN) lattice sites in $xy$ plane and neighboring layers are respectively represented by $t_1$ and $t_2$. $\gamma$ is the non-Hermitian strength and $\pm i\gamma$ denotes gain and loss at nonequivalent lattice sites labeled by different colors. All parameters are assumed to be real. In addition, the lattice constants both are unit either in $xy$ plane or along $z$ direction. For specific $k_z$ with nonzero values, the 3D layer stacking model can be reduced to a 2D singe-layer Haldane model with Hamiltonian:

$$H = \begin{pmatrix} i\gamma + t_2 f_2(k_x, k_y, \varphi) & t_1 f_1(k_x, k_y) \\ t_1 f_1(-k_x, k_y) & -i\gamma + t_2 f_2(k_x, k_y, -\varphi) \end{pmatrix} \quad (2)$$

where $f_1(k_x, k_y) = e^{ik_x/\sqrt{3}} + 2\cos(k_y/2) e^{-ik_x/2\sqrt{3}}$, $f_2(k_x, k_y, \varphi) = 2\cos(k_y - \varphi) + 4\cos(\sqrt{3}k_x/2)\cos(k_y/2 + \varphi)$. Phase factor $\varphi$ can be reviewed as the phase accumulation for $k_z$ in z direction of Eq. (1). The sign of $\varphi$ depends on the connecting direction of next-nearest-neighboring (NNN) lattice sites. Equations. (1) and (2) show that the 3D layer stacking model can be completely mapped to a 2D Haldane model. Their corresponding Brillouin zones in momentum space are displayed in Figs. 1(b) and 1(e), respectively. Capital letters are utilized to denote the highly symmetric points in the reciprocal space. It is notable that the valleys $K$ and $K'$ are inequivalent anymore because the chiral coupling (for 3D structure) or the complex NNN hopping (for 2D model) break the time-reversal symmetry. Therefore, the irreducible Brillouin zone in $xy$ plane turn to be the painted region in orange. Along the high symmetry line $\bar{H} - K - H$, bandgap of the 3D acoustic system would close at the value of $k_z = 0$ and $\pm\pi$ while other parameters are $t_1 = 1, t_2 = 1/3$ and $\gamma = 0.1$. For its 2D counterpart, if phase factor $\varphi$ takes zero value which is equivalent to the integer multiple of $\pi$ for $k_z$, then the coalesced points would also emerge at $K$ and $K'$ valleys as Fig. 1(f) shows. Otherwise, complete spectral gap occurs for nonzero phase factor with $\varphi = \pi/4$ for an example in Fig. 1(g). It indicates that the complex on-site potential can modulate the non-Hermitian topology of the Haldane model.

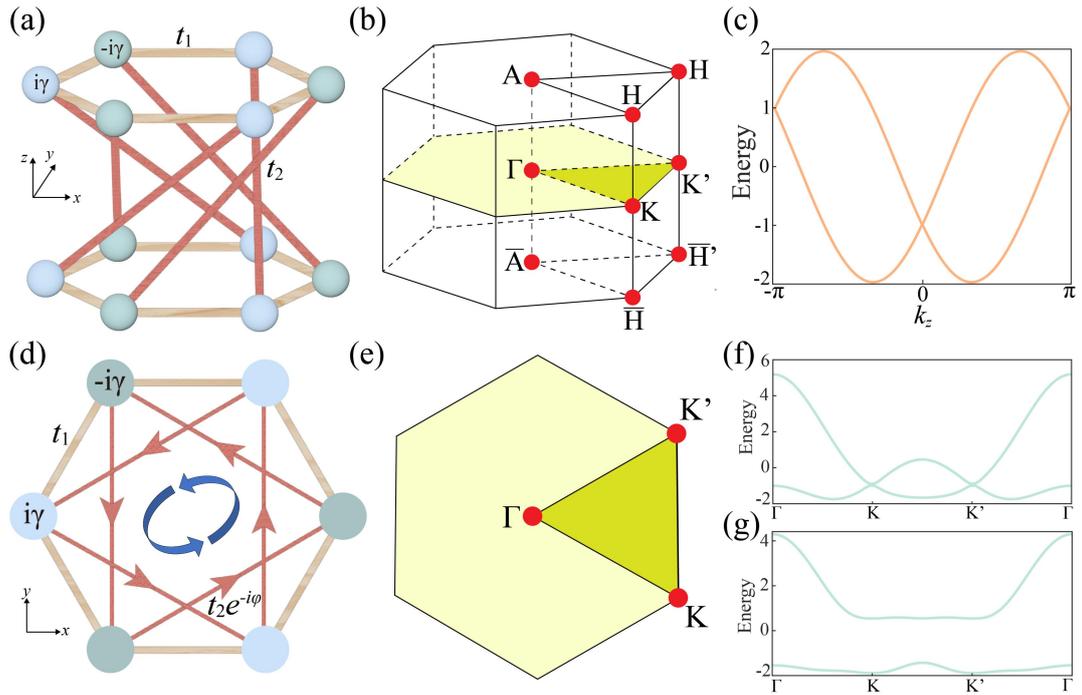

FIG. 1. (a) Sketch map of a layer stacking 3D model with chiral interlayer hopping and staggered gain and loss. (b) The Brillouin zone in momentum space corresponds to the model in (a). (c) The band structure as a function of $k_z$ along the high symmetry line $\bar{H} - K - H$ with parameters $t_1 = 1, t_2 = 1/3$ and $\gamma = 0.1$. (d) The projected map in $xy$ plane of the model in panel (a) with corresponding Brillouin zone in (e). (f) The dispersion relations along the contours of painted triangles in (e) when $t_1 = 1, t_2 = 1/3$, $\gamma = 0.1$ and $\varphi = 0$. If $\varphi$ is taken as $\pi/4$ and other parameters remain unchanged, the spectra evolve into (g).

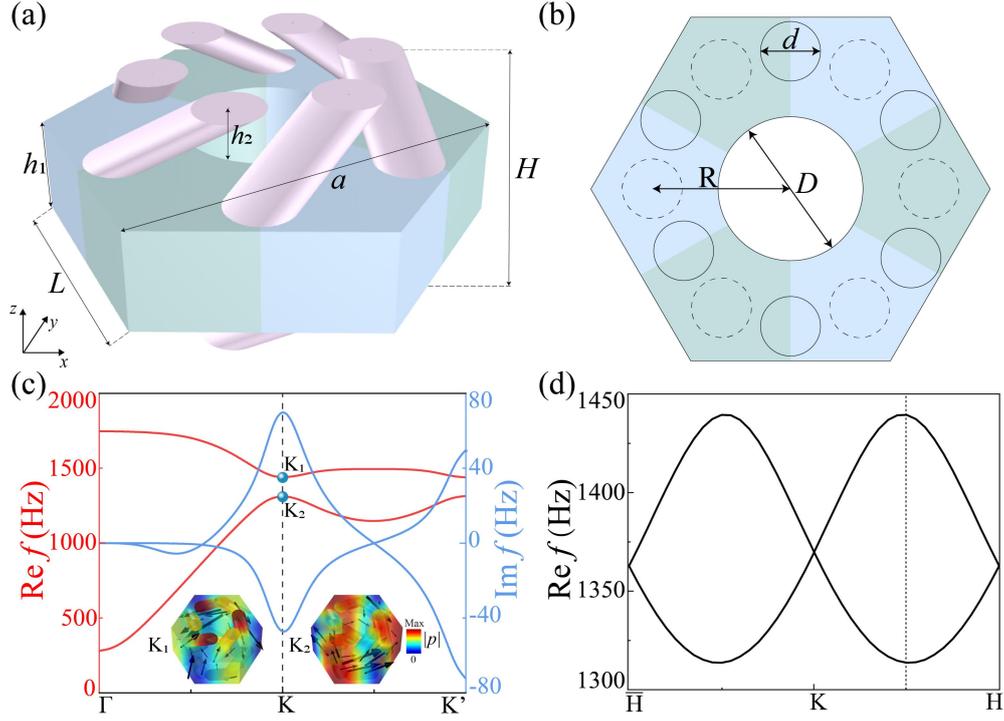

Fig. 2. (a) Acoustic realization of the tight-binding model of Fig. 1(a). (b) Top view of the 3D non-Hermitian SC. Solid and dotted circles denote the connecting position of tubes on the upper and lower surfaces of the cavities. (c) The real (red lines) and imaginary (blue lines) dispersion relations of the non-Hermitian SC when $k_z = \pi/2H$ and $\gamma = 0.1$. (d) Numerical calculation for the real part of dispersion relation along highly symmetric line $\bar{H} - K - H$ at $K$ position with dashed line marking the position of the maximum bandgap.

Inspired by the works [53, 56], in which effective gauge flux can be realized in a reduced 2D sonic system by engineering interlayer couplings when $k_z$ is a good quantum number. To realize the tight-binding model in Fig. 1(a), we designed a 3D acoustic crystal with Haldane model pattern in a single layer and chiral layer-stacking along $z$ direction as the side view depicts in Fig. 2(a). In addition, gain and loss are deliberately introduced on different types of lattice sites. It has been demonstrated that acoustic metamaterial is a versatile platform for the exploration of topological phenomena [57-60]. Here, a primitive unit cell is composed of a large hollow cavity and six chiral tubes connecting the upper and lower layers. Sound is encapsulated by the rigid boundaries of the cavities and tubes, which can be effectively viewed as sound hard boundaries. To introduce non-Hermiticity adequately, the air region inside the cavity is divided into six identical pieces centered at the corners. Three nonadjacent pieces are imposed gain by setting acoustic velocity of background air as $c = 343(1 + \gamma i)\ m/s$. The other three pieces are lossy regions by setting $c = 343(1 - \gamma i)\ m/s$. The coefficient $\gamma$ in the imaginary part of the velocities represents the non-Hermitian strength. And the density of air is $\rho = 1.23\ kg/m^3$. The side length and height of the hexagonal cavity is $L = h_1 = 8cm$. Then the lattice constant in $xy$ plane is $a = \sqrt{3}L = 8\sqrt{3}cm$. In $z$ direction, interlayer distance is $2h_2 = 5cm$. Hence, the lattice constant in this direction is $H = h_1 + 2h_2 = 13cm$. In Fig. 2(b), we also give the top view of a unit cell of the sonic system. It shows that each concentric cavity has a hole with diameter $D = 6cm$ inside the central region. Interlayer chiral tubes host diameter $d = 2.4cm$, and the distance of their connecting position from the center of cavities is $R = 5.5cm$. For a fixed $k_z$, the interlayer couplings in 3D structure can be viewed as the NNN coupling in reduced 2D systems and the hopping phase $\phi$ is accumulated by the acoustic propagation along $z$ direction $\phi = k_z H_z$. And the sign of hopping phase $\phi$ is determined by the chiral coupling direction. More specifically, anticlockwise and clockwise chiral coupling respectively accumulate negative and positive hopping phase $\phi$. Here, only anticlockwise interlayer coupling is considered. For the selected parameters given above, the real and imaginary dispersion relations of the non-Hermitian SC can be numerically calculated by commercial software COMSOL Multiphysics based on finite element method when Bloch periodic boundary conditions are imposed on the outer surfaces. Distinct from the Hermitian SC, the eigenfrequencies of the non-Hermitian system are complex. And Fig. 2(c) presents the 2D band diagram for the real (red lines) and imaginary (blue lines) dispersion relations at $k_z = \pi/2H$. A direct band gap of the real frequencies emerges at high symmetry point $K$ in reciprocal space. The acoustic amplitude (color) and energy flow (arrows) are also displayed in the insets. At extreme value points $K_1$ and $K_2$, the acoustic energies concentrate on different valleys and flow toward opposite directions. The valley provides an alternative freedom to control the acoustic propagation and confinement. The first-principles simulation of the band structure in Fig. 2(d) manifests that topological transition happens when $k_z$ is detuned along the hinge at the valley. It is worth noting that the maximal complete band gap along highly symmetric line $\bar{H} - K - H$ just appears at $k_z = \pi/2H$ at highly symmetric point $K$ as Fig. 2(d) shows. And $k_z = 0$ is the phase transition position. The simulated result is consistent with analytical one in Fig. 1(c). Then throughout the whole paper, $k_z = \pi/2H$ is fixed unless mentioned otherwise.

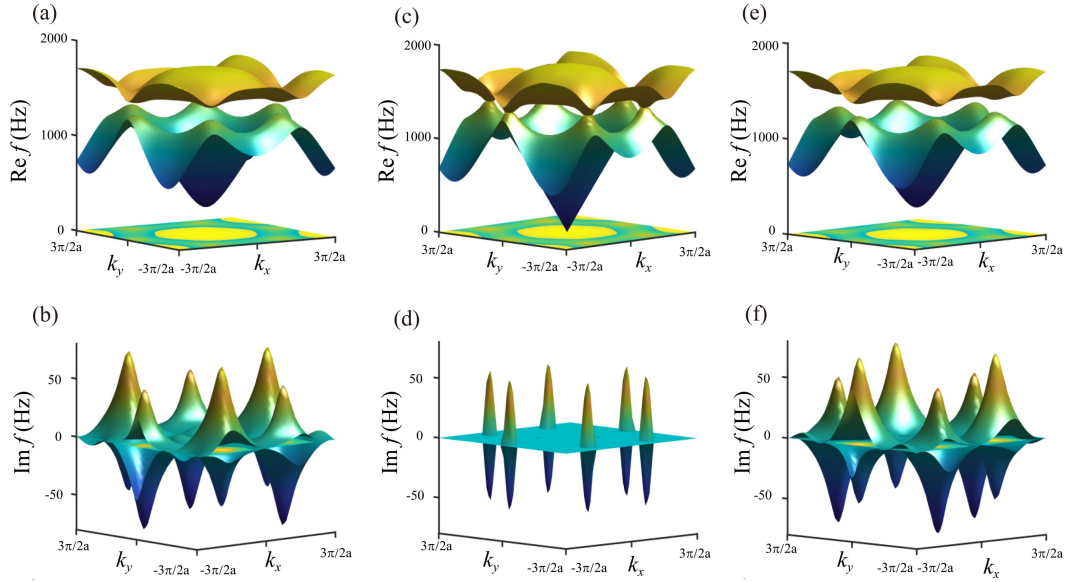

FIG. 3. The real (a) and imaginary (b) energies of the 3D band structures at $k_z = -\pi/2H$. Three pairs of Dirac points are completely gapped in the real spectrum. At $k_z = 0$, Dirac points turn into Dirac rings in both the real (c) and imaginary (d) energy plots as the introduction of non-Hermiticity. In the cases with positive $k_z$ values, Dirac ring would be separated by complete gap in real spectra with $k_z = \pi/2H$ for an example in (e). (f) The 3D imaginary spectrum diagram at $k_z = \pi/2H$.

To better understand the topological transition, we study 3D band structures at several discrete values $k_z = -\pi/2H$, 0 and $\pi/2H$. It is well known that either broken time-reversal symmetry or inversion symmetry can gap out the Dirac points. For $k_z \neq 0$, band gap originates from broken time-reversal symmetry in the designed acoustic metamaterial as directional hopping phase is introduced by the chiral coupling between layers stacked along $z$ direction. In Figs. 3(a) and 3(b), we show the real and imaginary eigenfrequencies at $k_z = -\pi/2H$. In comparison, the corresponding spectra at $k_z = \pi/2H$ are also plotted and displayed in Figs. 3(e) and 3(f). We can indeed see that the Dirac points are completely gapped in their real spectra. And acoustic valleys emerge at highly symmetric points $K$ and $K'$. However, due to the emergence of pseudo-inversion symmetry with $c(\vec{r}) = c^+(-\vec{r})$ and $\rho(\vec{r}) = \rho^+(-\vec{r})$, Dirac points at $K$ and $K'$ valleys cannot be destroyed and evolve into Dirac rings at $k_z = 0$ as Figs. 3(c) and 3(d) exhibit. The band diagrams of the 3D acoustic metamaterial with opposite $k_z$ values are the same, however their Berry curvatures and valley Chern numbers are exactly opposite. The signs of the valley Chern numbers are tightly related with that of $k_z$. The Berry curvature of the $n$th band is defined as $\Omega_n(\boldsymbol{k}) = \nabla_{\boldsymbol{k}} \times A_n(\boldsymbol{k})$ where $A_n(\boldsymbol{k}) = i\langle u_{n,\boldsymbol{k}}^L | \nabla_{\boldsymbol{k}} | u_{n,\boldsymbol{k}}^R \rangle$ is the Berry connection and $u_{n,\boldsymbol{k}}$ is the Bloch state. The valley Chern number of the band is defined as $C_v = C_k - C_{k'}$. And the valley Chern numbers at different valleys can be achieved by integration of Berry curvature. In sharp contrast to the Hermitian case, the right and left eigenvectors of the non-Hermitian Hamiltonian are in general different and not necessarily mutually orthogonal. The biorthogonal set is achieved by the normalization method $\langle u_i^L | u_j^R \rangle = \delta_{ij}$. Overall, the non-Hermitian structures sharing the same dispersion relations at opposite $k_z$ can have analogous topologies characterized by inverse topological invariants. Therefore, in the following only $k_z$ values larger than zero are focused on, and results can be similarly attained for those smaller than zero.

## III. ANALYSIS OF GAIN AND LOSS SURFACE STATES

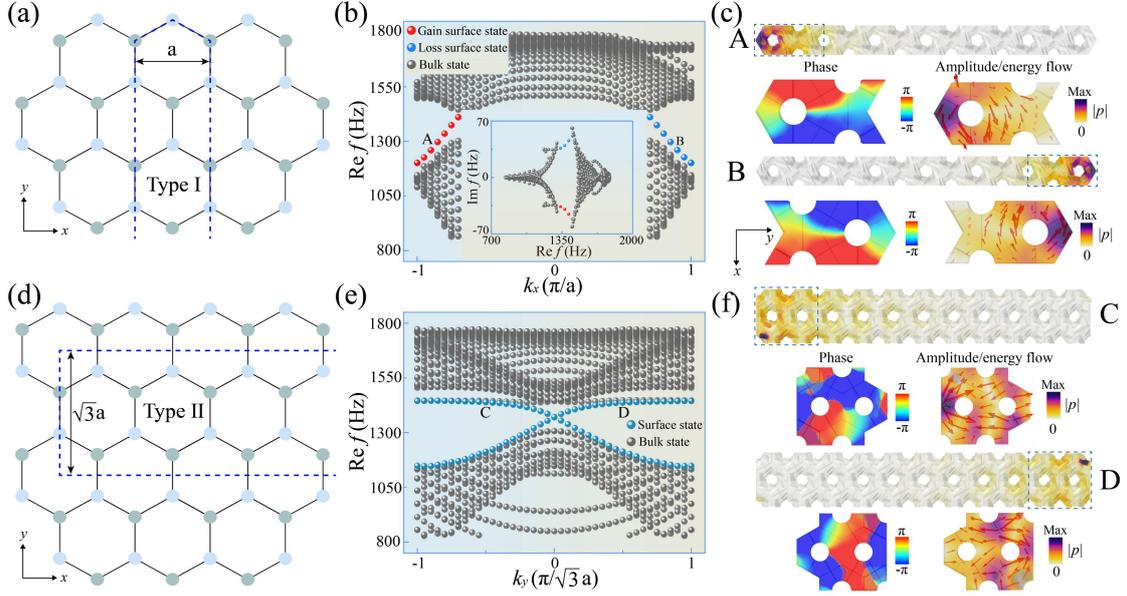

FIG. 4. (a) Schematics of the non-Hermitian sonic system with zigzag boundaries in $x$ direction and armchair in $y$ direction. A ribbon-shaped supercell with zigzag boundaries termed Type I structure is outlined by blue dashed lines. (b) The real projected bands for the supercell in (a). Gain and loss surface states are denoted by red and blue dots. The inset shows the complex energy spectrum of projected bands. (c) Plots of the acoustic pressure amplitude, phase and Poynting vector for the eigenstates at A and B points of the surface dispersions. The counterparts in Type II supercell with armchair boundaries are displayed in (d) - (f), respectively. Other parameters are chosen as $k_z = \pi/2H$ and $\gamma = 0.1$.

It is known that in graphene, edge modes within different regimes of reciprocal space bound at distinct boundaries [61]. Here we numerically demonstrate that surface dispersions with both of amplified and attenuated branches can be produced in a non-Hermitian sample with zigzag terminations if gain and loss are included. Due to the emergent periodicity in z direction, edge and corner states are in fact surface and hinge states. As the schematics in Fig. 4(a) visually shows, the 3D acoustic sample host zigzag edges in $x$ direction while armchair along $y$ direction. A portion of the structure outlined by blue dashed lines with zigzag boundaries in $x$ direction is termed as Type I ribbon-shaped supercell. Bloch periodic boundary conditions are exerted on $x$ and z direction and its projected band structure are numerically calculated and exhibited in Fig. 4(b). In the real diagram, we find that except for bulk states, there also exists two branches of surface dispersions connecting the upper and lower bulk spectra. One branch is referred to as gain surface states (red dots) as their imaginary frequencies are negative, while the other branch (blue dots) with positive imaginary frequencies denotes loss surface modes as the complex energy spectra in the inset shows. It unambiguously indicates that surface modes exhibit completely different behaviors from the Hermitian analog when gain and loss are included in the SC. Visualization of wavefunctions shows that the phase profiles of the acoustic field at two opposite wavevectors carry opposite phase vortices and winding directions [see Fig. 4(c)]. It indicates that two eigenstates with opposite wavevectors carry opposite angular momentum. In addition, the profiles of the acoustic Poynting vectors also named energy flows (indicated by red arrows) manifest the same point. But the acoustic intensity confine at opposite sides of the supercell. The lower and upper zigzag boundaries respectively support gain and loss surface modes. This intriguing feature that different boundaries sustain distinct branches of edge modes is similar to that in quantum spin Hall effect [62, 63]. For comparison, we also consider other type supercell named Type II as the dashed lines depict in Fig. 4(d). In this case, the real projected bands are calculated and exhibited in Fig. 4(e). The acoustic profiles in Fig. 4(f) reveal that edge modes at inverse wavevectors also accumulate at opposite endings. However, chaotic arrows imply that the acoustic energy flow toward all directions and cannot form effective edge current. Then it can be reasonably expected that armchair boundaries in this sample unable to support gain/loss surface transport.

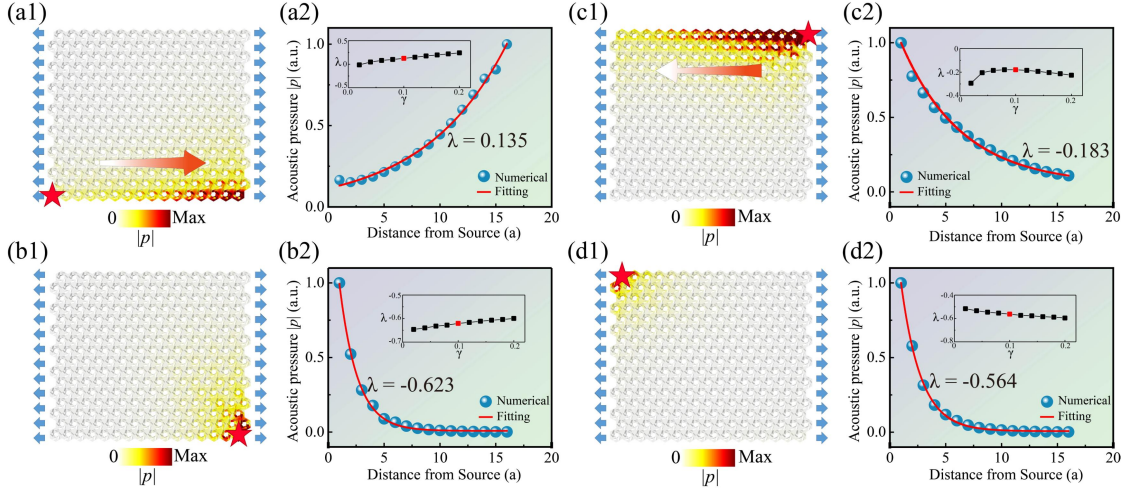

FIG. 5. Demonstration of directional transport for gain and loss surface states. Profiles of acoustic pressure fields and their quantitative plots at the sample boundaries when sound source is positioned at lower left hinge (a1- a2), lower right hinge (b1-b2), upper right hinge (c1-c2) and upper left hinge (d1-d2). Excited sources are denoted by red stars. The left and right sides of the rectangular samples are terminated by armchair boundaries and set as open boundaries and the upper and lower sides are zigzag boundaries. Panels (a1) - (d1) visualize the eigenfunctions of the acoustic field and Panels (a2) - (d2) present the quantitative description of the transmitting sound along the propagating path. The excited frequency is selected as $f = 1350$Hz at $k_z = \pi/2H$ with non-Hermitian strength $\gamma = 0.1$.

To perceptually verify gain and loss surface states we mentioned above, we design a rectangular sample with armchair shapes on the left and right sides while zigzag-shaped terminations on the upper and lower. Non-Hermitian surface states only survive at zigzag boundaries, then armchair boundaries are set as open boundary conditions denoting by blue arrows in Figs. 5(a1) - (d1). A point source with excited frequency $f = 1350$Hz is placed at lower-left hinge of the rectangular sample. Only one branch of surface states with positive group velocity will be excited, and the propagating mode is exhibited in Fig. 5(a1). This concludes that the surface dispersion supporting positive group velocity are indeed gain surface states. At the same boundary, when source is placed at the lower-right hinge, sound would stay still because of the absence of surface band with negative wavevector [Fig. 5(b1)]. Similar analysis can also be performed on the transport at the upper boundary, which supports left-propagating surface states. Similarly, when sound source is placed at the right ending of the upper boundary, the decreasing sound field in Fig. 5(c1) unequivocally indicates that the excited surface mode is lossy. If excited place changes to left side, no surface modes can be stimulated [Figs. 5(d1)]. Furthermore, the non-Hermitian surface states can be analyzed quantitatively. The acoustic amplitude within each unit cell at the excited boundaries is summed and presented by blue dots, as Figs. 5(a2) - 5(d2) shows. The numerically calculated results can be fitted by the exponential function $|\psi_n| \propto e^{\lambda n}$ (red lines), in which $|\psi_n|$ denotes the acoustic amplitude at the $n$th site of the zigzag boundaries and coefficient $\lambda$ represents amplified (positive) or attenuated (negative) strength. For the four excited scenarios, the fitting coefficients $\lambda$ are retrieved and given in the corresponding panels. In addition, we also consider the effect of non-Hermitian strength on the transport of dissipative surface modes. And fitting coefficient $\lambda$ as a function of non-Hermitian strength $\gamma$ is plotted in the insets of Figs. 5(a2) - (d2). The overall trend in the fitting curves reveals that greater non-Hermitian strength results in larger amplification or attenuation, indicating that the surface modes become intense or weak more quickly.

From the analysis in Section II, we identify that the interlayer couplings introduce hopping phase between NNN sites, leading to the emergence of Chern insulator for fixed $k_z$. In conventional acoustic Chern insulators, topologically protected surface states enable backscattering suppression [64]. However, with the incorporation of on-site gain and loss, acoustic chiral current is generated at the zigzag boundary as Fig. 4(a) depicts. And the upper and lower zigzag boundaries share artificial magnetic field with the same direction, thereby resulting in gain surface states at the lower boundary and loss modes at the upper boundary. In particular, one horizontal boundary supports amplified mode and the other sustains damped mode. This is the underlying mechanism for gain and loss surface states.

## IV. HYBRID SKIN-TOPOLOGICAL EFFECTS IN TUNABLE SAMPLES AND ITS ROBUSTNESS

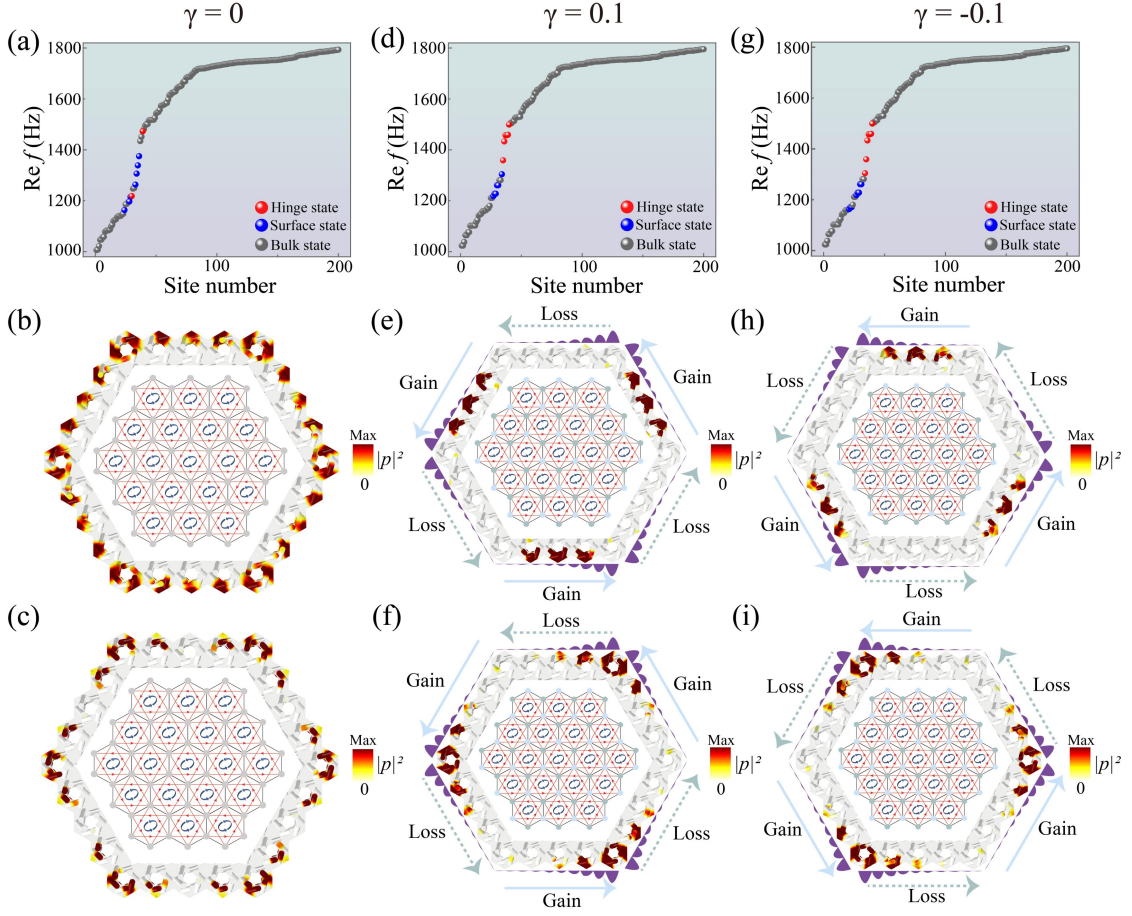

FIG. 6. Numerical simulations of hinge skin states originating from gain and loss surface states in a hexagonal sample. (a) The real energy spectra; Field maps for (b) chiral surface mode and (c) hinge state in a lossless acoustic metamaterial. If dissipation is considered and the non-Hermitian strength is selected as $\gamma = 0.1$, real spectra is calculated and exhibited in (d). Except for the surface states (e), hinge skin states also emerge (f). For $\gamma = -0.1$, the spectra in (g) are the same as that in (e), but the residing positions of surface states and hinge skin modes shift to the other three boundaries (h) and hinges (i). Red, blue and gray dots in the spectra diagram denote hinge, surface and bulk states, respectively. The insets in the profiles aim to show the detail of the sample boundaries.

Different from traditional mechanisms to generate corner states such as

Jackiw-Rebbi mechanism [65] and edge bands with distinct Zak phases [16]. In this work, hinge states can be precisely predicted if gain and loss surface states are alternatively positioned at two adjoint zigzag edges. To be more specific, when gain and loss surface modes propagate toward the same direction on two adjacent boundaries, then their superimposed modes would come into being at their intersected position, which are termed as hinge skin states. And the hinge skin modes could also be expressed as $\psi \propto c_1 e^{\lambda_1 n} + c_2 e^{\lambda_2 n}$, in which one term represents gain mode and the other denotes loss mode. Hinge skin mode then can also be viewed as the superposition of two non-Hermitian surface modes.

In the following, we judiciously design acoustic arrays with different non-Hermiticities at boundaries to demonstrate the consequence of gain and loss edges on the hinge skin states. In Fig. 6, a hexagonal array with 61 primitive units is constructed. For Hermitian limit, i.e., non-Hermitian strength $\gamma = 0$, the real spectrum comprises bulk states, surface states and hinge states [Fig. 6(a)]. The wavefunctions of a randomly selective surface state and hinge state are numerically attained in Figs. 6(b) and 6(c). The color denotes the sound intensity and the abbreviated insets inside the field map show the outlines of the samples. For $\gamma = 0.1$, the hexagonal array hosts three alternative gain and loss boundaries. Owing to nonzero artificial gauge fluxes on boundary plaquettes, chiral surface currents subsequently arise. Normal surface states thus evolve into gain and loss modes [denoted by blue dots in Fig. 6(d)], resulting in hinge skin states [red dots in Fig. 6(d)]. The field maps of a selected non-Hermitian surface state and hinge skin mode are respectively presented in Figs. 6(e) and 6(f). The arrows schematically show the directional transport of gain and loss surface modes with wavy curves indicating the exponential decay or enhancement of sound intensity. If non-Hermitian strength is reversed to $\gamma = -0.1$, the real spectra sustain as the swapping of gain and loss do not affect the eigen energies [Fig. 6(g)]. However, the confined positions of non-Hermitian surface modes [Fig. 6(h)] and hinge skin states [Fig. 6(i)] would shift to the other three nonadjacent surfaces or hinges.

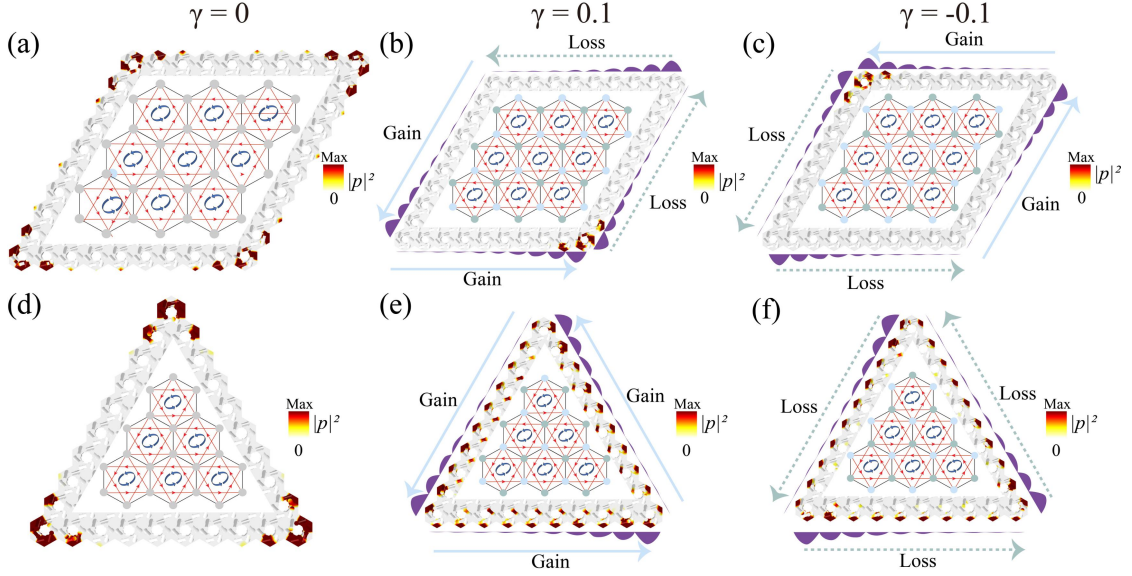

FIG. 7. Numerical calculation of localized states in rhombic and triangular samples for diverse dissipative strength. The profiles of wavefunctions are displayed in (a)(d) for $\gamma = 0$, in (b)(e) for $\gamma = 0.1$ and in (c)(f) for $\gamma = -0.1$. The first row demonstrates the hinge states in rhomb-shaped array (a)-(c), while the simulated results of triangular-shaped array in the second row are for hinge states (d) and surface states (e) - (f).

As expected from the above discussion, the non-Hermitian-controlled topological boundaries can be flexibly reconfigured in different samples to steer sound. To demonstrate such versatile topological sound steering, the array is switched from hexagonal to rhombic and triangular. In rhomb-shaped array with four zigzag boundaries, hinge states accumulate at four hinges attributing to the higher-order topological effect as Fig. 7(a) shows. If complex on-site potentials are included according to Haldane model pattern, two successive gain and loss surfaces would meet at two opposite hinges. Because synthetic gauge flux is anticlockwise as the arrows show in the insets. Sound transmitted along the same direction would be trapped at the joint position from gain to loss surfaces. Hinge skin states thus can be predicted at the lower-right hinge if $\gamma = 0.1$. And theoretical analysis can be evidenced by the simulated result in Fig. 7(b). Similarly, skin modes can also be expected at the upper-left hinge for $\gamma = -0.1$, as the simulation in Fig. 7(c) shows. If the acoustic array is reconfigured as triangular, there are hinge states at $\gamma = 0$ [Fig. 7(d)]. In comparison, in non-Hermitian case all the boundaries become gain or loss. And hinge skin states are necessarily impossible to appear due to the absence of concurrent gain and loss surfaces. Localized states in non-Hermitian cases only present as gain hinge modes [Fig. 7(e)] or loss hinge modes [Fig. 7(f)] for $\gamma = 0.1$ and $-0.1$, respectively.

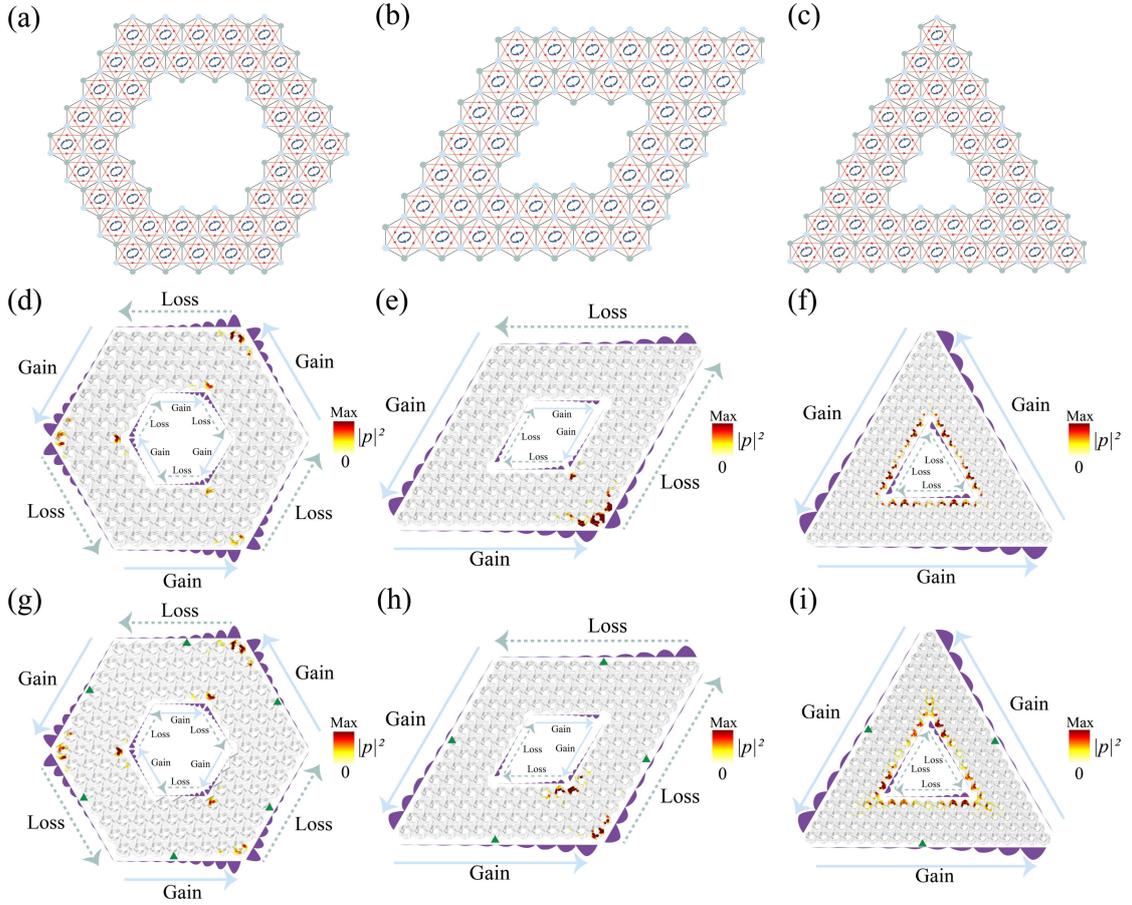

FIG. 8. (a) - (c) Schematic sketch of hexagonal, rhombic and triangular samples with concentric holes sharing similar shape to the samples. For $\gamma = 0.1$, hinge states induced by hybrid skin-topological effect both take place in the interior or/and exterior surfaces of hexagonal sample (d) and rhombic array (e), while the bound states present as surface states in triangular array (f). If complex potential at one selected position of each edge labeled by small green triangles is removed, the corresponding profile at the same frequency still preserve as panels (g) - (i) exhibit, demonstrating the robustness of non-Hermitian boundaries to defects.

Based on the above exploration of hinge skin states, we assume that versatile platform supporting multiple hinge skin modes can always be accessed by intentionally expanding the lattice domain and cutting the surfaces in specific pattern to construct more surfaces to support multiple hinge skin modes. However, large lattice is bulky and costly to fabricate in experiment, which will restrict the construction of versatile acoustic samples. In this proposal, the hinge skin states are mostly built at the boundaries and scarcely rest on the bulk of acoustic metamaterial. Therefore, one idea can be conceived to figure out this issue through cutting the central area of the arrays. Without loss of generality, we remove the central regions in hexagonal, rhombic and triangular samples as the sketches in Figs. 8(a) - (c) depict. The central holes share the same geometry with their outer surfaces. This guarantees the internal and external surfaces the feasibility to support hinge skin states to the maximum extent. The gain surfaces at outer regions correspond to the loss surfaces at inner domains and vice versa. In addition, the gauge flux directions along the internal and external surfaces of the non-Hermitian SCs are also reversed. Overall, the hinges supporting skin modes are at the same angles for both surfaces. The analysis can be precisely validated by the simulated results in Figs. 8(d) and 8(e). In triangular acoustic arrays, the inner surfaces support non-Hermitian surface modes as their outer ones as demonstrated in Fig. 8(f). At last, we demonstrate the robustness of the acoustic arrays to defects. Robustness is one of the most intriguing features for conventional topological edge states, which can substantially reduce the backscattering loss and thus has inspired a variety of applications in integrated optics [66] and energy harvesting [9], as well as directional transport with high capacity [67, 68]. Our non-Hermitian-controlled sound trapping scheme is also topologically robust against defects. Even though defects are intentionally created along the structural surfaces by eliminating the complex on-site potentials at lattice sites of surfaces. The gain and loss modes can still detour around the defect and form effective hinge skin states, as shown in Figs. 8(g) - (i). Excellent agreement can be achieved for the result in both cases with and without defects.

## V. CONCLUSION AND OUTLOOK

In conclusion, we have successfully demonstrated hybrid skin-topological effect in macroscopic SC by judiciously introducing complex on-site potential and artificial gauge field. Our proposed scheme is quite practical to realize in acoustic experiments and other experimental platforms. Similar effects can also be achieved with alternative designs featuring specific non-Hermitian configurations. They could also be adapted to higher frequencies from audible to ultrasonic frequencies by appropriately scaling down the lattice constant. Tunable higher-order skin-topological hinge modes are beneficial for constructing integrated devices in sound. Many applications such as signal capture and obstacle orientation may benefit from the robustness to disorders or immunity to backscattering in the designed devices.

## ACKNOWLEDGMENTS

We are grateful to Dr. Weiwei Zhu for helpful discussions. The authors acknowledge support from the National Natural Science Foundation of China (NNSFC) (Grant No. 12047541, No. 12404493 and No. 12274077) and the Research Fund of Guangdong-Hong Kong-Macao Joint Laboratory for Intelligent Micro-Nano Optoelectronic Technology (Grant No. 2020B1212030010).

**References:**
[1] M. Z. Ke, Z. Y. Liu, C. Y. Qiu, W. G. Wang, J. Shi, W. J. Wen, and P. Sheng, Negative-refraction imaging with two-dimensional phononic crystals, Phys. Rev. B **72**, 064306 (2005).
[2] Y. Wu, and J. Li, Total reflection and cloaking by zero index metamaterials loaded with rectangular dielectric defects, Appl. Phys. Lett. **102**, 183105 (2013).
[3] L. Zheng, Y. Wu, X. Ni, Z. Chen, M. Lu, and Y. Chen, Acoustic cloaking by a near-zero-index phononic crystal, Appl. Phys. Lett. **104**, 161904 (2014).
[4] R. Fleury, D. L. Sounas, C. F. Sieck, M. R. Haberman, and A. Alu, Sound Isolation and Giant

Linear Nonreciprocity in a Compact Acoustic Circulator, Science **343**, 516 (2014).

[5] X. Ni, C. He, X. Sun, M. Lu, and Y. Chen, Topologically protected one-way edge mode in networks of acoustic resonators with circulating air flow, New J. Phys. **17**, 053016 (2015).

[6] Z. Yang, F. Gao, X. Shi, X. Lin, Z. Gao, Y. Chong, and B. Zhang, Topological Acoustics, Phys. Rev. Lett. **114**, 114301 (2015).

[7] C. He, X. Ni, H. Ge, X. Sun, Y. Chen, M. Lu, X. Liu, and Y. Chen, Acoustic topological insulator and robust one-way sound transport, Nat. Phys. **12**, 1124 (2016).

[8] Z. Zhang, Q. Wei, Y. Cheng, T. Zhang, D. Wu, and X. Liu, Topological Creation of Acoustic Pseudospin Multipoles in a Flow-Free Symmetry-Broken Metamaterial Lattice, Phys. Rev. Lett. **118,** 084303 (2017).

[9] S. Wu, Y. Wu, and J. Mei, Topological helical edge states in water waves over a topographical bottom, New J. Phys. **20**, 023051 (2018).

[10] J. Lu, C. Qiu, M. Ke, and Z. Liu, Valley Vortex States in Sonic Crystals, Phys. Rev. Lett. **116**, 093901 (2016).

[11] J. Lu, C. Qiu, L. Ye, X. Fan, M. Ke, F. Zhang, and Z. Liu, Observation of topological valley transport of sound in sonic crystals, Nat. Phys. **13**, 369 (2017).

[12] J. Lu, C. Qiu, W. Deng, X. Huang, F. Li, F. Zhang, S. Chen, and Z. Liu, Valley Topological Phases in Bilayer Sonic Crystals, Phys. Rev. Lett. **120**, 116802 (2018).

[13] X. Ni, M. Weiner, A. Alù, and A. B. Khanikaev, Observation of higher-order topological acoustic states protected by generalized chiral symmetry, Nat. Mater. **18**, 113 (2018).

[14] H. Fan, B. Xia, L. Tong, S. Zheng, and D. Yu, Elastic Higher-Order Topological Insulator with Topologically Protected Corner States, Phys. Rev. Lett. **122**, 204301 (2019).

[15] H. Xue, Y. Yang, G. Liu, F. Gao, Y. Chong, and B. Zhang, Realization of an Acoustic Third-Order Topological Insulator, Phys. Rev. Lett. **122**, 244301 (2019).

[16] X. Zhang, H. Wang, Z. Lin, Y. Tian, B. Xie, M. Lu, Y. Chen, and J. Jiang, Second-order topology and multidimensional topological transitions in sonic crystals, Nat. Phys. **15**, 582 (2019).

[17] S. Wu, B. Jiang, Y. Liu, and J. Jiang, All-dielectric photonic crystal with unconventional higher-order topology, Photonics Res. **9**, 668 (2021).

[18] S. Wu, Z. Lin, B. Jiang, X. Zhou, Z. H. Hang, B. Hou, and J. Jiang, Higher-order Topological States in Acoustic Twisted Moiré Superlattices, Phys. Rev. Appl. **17**, 034061 (2022).

[19] S. Wu, Z. Lin, Z. Xiong, B. Jiang, and J. Jiang, Square-Root Higher-Order Topology in Rectangular- Lattice Acoustic Metamaterials, Phys. Rev. Appl. **19**, 024023 (2023).

[20] H. Wang, Z. Lin, B. Jiang, G. Guo, and J. Jiang, Higher-Order Weyl Semimetals, Phys. Rev. Lett. **125**, 146401 (2020).

[21] M. Xiao, L. Ye, C. Qiu, H. He, Z. Liu, and S. Fan, Experimental demonstration of acoustic semimetal with topologically charged nodal surface, Sci. Adv. **6**, eaav2360 (2020).

[22] L. Luo, H. Wang, Z. Lin, B. Jiang, Y. Wu, F. Li, and J. Jiang, Observation of a phononic higher-order Weyl semimetal, Nat. Mater. **20**, 794 (2021).

[23] Q. Wei, X. Zhang, W. Deng, J. Lu, X. Huang, M. Yan, G. Chen, Z. Liu, and S. Jia, Higher-order topological semimetal in acoustic crystals, Nat. Mater. **20**, 812 (2021).

[24] S. Yao, and Z. Wang, Edge States and Topological Invariants of Non-Hermitian Systems, Phys. Rev. Lett. **121**, 086803 (2018).

[25] L. Jin, and Z. Song, Bulk-boundary correspondence in a non-Hermitian system in one dimension with chiral inversion symmetry, Phys. Rev. B **99**, 026808 (2019).

[26] C. H. Lee, and R. Thomale, Anatomy of skin modes and topology in non-Hermitian systems, Phys. Rev. B **99**, 201103 (2019).

[27] L. Li, C. H. Lee, S. Mu, and J. Gong, Critical non-Hermitian skin effect, Nat. Commun. **11**, 5491 (2020).

[28] L. Zhang, Y. Yang, Y. Ge, Y. J. Guan, Q. Chen, Q. Yan, F. Chen, R. Xi, Y. Li, D. Jia, S. Q. Yuan, H. X. Sun, H. Chen, and B. Zhang, Acoustic non-Hermitian skin effect from twisted winding topology, Nat. Commun. **12**, 6297 (2021).

[29] W. Zhu, W. X. Teo, L. Li, and J. Gong, Delocalization of topological edge states, Phys. Rev. B **103**, 195414 (2021).

[30] Y. Qin, K. Zhang, and L. Li, Geometry-dependent skin effect and anisotropic Bloch oscillations in a non-Hermitian optical lattice, Phys. Rev. A **109**, 023317 (2024).

[31] X. Zhang, Y. Tian, J. Jiang, M. Lu, and Y. Chen, Observation of higher-order non-Hermitian skin effect, Nat. Commun. **12**, 5377 (2021).

[32] S. Liu, R. Shao, S. Ma, L. Zhang, O. You, H. Wu, Y. J. Xiang, T. J. Cui, and S. Zhang, Non-Hermitian Skin Effect in a Non-Hermitian Electrical Circuit, Research **2021**, 5608038 (2021).

[33] K. Takata, and M. Notomi, Photonic Topological Insulating Phase Induced Solely by Gain and Loss, Phys. Rev. Lett. **121**, 213902 (2018).

[34] H. Gao, H. Xue, Q. Wang, Z. Gu, T. Liu, J. Zhu, and B. Zhang, Observation of topological edge states induced solely by non-Hermiticity in an acoustic crystal, Phys. Rev. B **101**, 180303 (2020).

[35] Y. Wu, C. Liu, and J. Hou, Wannier-type photonic higher-order topological corner states induced solely by gain and loss, Phys. Rev. A **101**, 043833 (2020).

[36] A. Guo, G. J. Salamo, D. Duchesne, R. Morandotti, M. Volatier-Ravat, V. Aimez, G. A. Siviloglou, and D. N. Christodoulides, Observation of P T -Symmetry Breaking in Complex Optical Potentials, Phys. Rev. Lett. **103**, 093902 (2009).

[37] D. Kip, C. E. Rüter, K. G. Makris, R. El-Ganainy, D. N. Christodoulides, and M. Segev, Observation of parity–time symmetry in optics, Nat. Phys. **6**, 192 (2010).

[38] W. Song, W. Sun, C. Chen, Q. Song, S. Xiao, S. Zhu, and T. Li, Breakup and Recovery of Topological Zero Modes in Finite Non-Hermitian Optical Lattices, Phys. Rev. Lett. **123**, 165701 (2019).

[39] Z. Xu, R. Zhang, S. Chen, L. Fu, and Y. Zhang, Fate of zero modes in a finite Su-Schrieffer-Heeger model with PT symmetry, Phys. Rev. A **101**, 013635 (2020).

[40] H. Gao, H. Xue, Z. Gu, T. Liu, J. Zhu, and B. Zhang, Non-Hermitian route to higher-order topology in an acoustic crystal, Nat. Commun. **12**, 1888 (2021).

[41] R. Okugawa, R. Takahashi, and K. Yokomizo, Second-order topological non-Hermitian skin effects, Phys. Rev. B **102**, 241202(R) (2020).

[42] K. Kawabata, M. Sato, and K. Shiozaki, Higher-order non-Hermitian skin effect, Phys. Rev. B. **102**, 205118 (2020).

[43] C. H. Lee, L. Li, and J. Gong, Hybrid Higher-Order Skin-Topological Modes in Nonreciprocal Systems, Phys. Rev. Lett. **123**, 016805 (2019).

[44] W. Zhu, and J. Gong, Hybrid skin-topological modes without asymmetric couplings, Phys. Rev. B **106**, 035425 (2022).

[45] Y. Li, C. Liang, C. Wang, C. Lu, and Y. Liu, Gain-Loss-Induced Hybrid Skin-Topological Effect, Phys. Rev. Lett. **128**, 223903 (2022).

[46] W. Zhu, and J. Gong, Photonic corner skin modes in non-Hermitian photonic crystals, Phys. Rev. B **108**, 035406 (2023)

[47] M. Wei, Y. Wang, M. Liao, Y. Yang, and J. Xu, Nonlinear topological laser on the non-Hermitian Haldane model with higher-order corner states, Opt. Express **31**, 39424 (2023).

[48] Y. Sun, X. Hou, T. Wan, F. Wang, S. Zhu, Z. Ruan, and Z. Yang, Photonic Floquet Skin-Topological Effect, Phys. Rev. Lett. **132**, 063804 (2024).

[49] Y. Zhang, L. Wang, Y. Liu, Z. Chen, and J. Jiang, Hybrid skin-topological effect in non-Hermitian checkerboard lattices with large Chern numbers, arXiv:2411.07465 (2024).

[50] J. Wu, R. Zheng, J. Liang, M. Ke, J. Lu, W. Deng, X. Huang, and Z. Liu, Spin-Dependent Localization of Helical Edge States in a Non-Hermitian Phononic Crystal, Phys. Rev. Lett. **133**, 126601 (2024).

[51] R. Zheng, J. Lin, J. Liang, K. Ding, J. Lu, W. Deng, M. Ke, X. Huang, and Z. Liu, Experimental probe of point gap topology from non-Hermitian Fermi-arcs, Commun. Phys. **7**, 298 (2024).

[52] L. Lu, L. Fu, J. D. Joannopoulos, and M. Soljačić, Weyl points and line nodes in gyroid photonic crystals, Nat. Photonics **7**, 294 (2013).

[53] M. Xiao, W. Chen, W. He, and C. T. Chan, Synthetic gauge flux and Weyl points in acoustic systems, Nat. Phys. **11**, 920 (2015).

[54] W. Chen, M. Xiao, and C. T. Chan, Photonic crystals possessing multiple Weyl points and the experimental observation of robust surface states, Nat. Commun. **7**, 13038 (2016).

[55] F. Li, X. Huang, J. Lu, J. Ma, and Z. Liu, Weyl points and Fermi arcs in a chiral phononic crystal, Nat. Phys. **14**, 30 (2017).

[56] C. He, H. S. Lai, B. He, S. Y. Yu, X. Xu, M. H. Lu, and Y. F. Chen, Acoustic analogues of three-dimensional topological insulators, Nat. Commun. **11**, 2318 (2020).

[57] Z. Yang, F. Gao, Y. Yang, and B. Zhang, Strain-Induced Gauge Field and Landau Levels in Acoustic Structures, Phys. Rev. Lett. **118**, 194301 (2017).

[58] Q. Wang, Y. Ge, H. Sun, H. Xue, D. Jia, Y. Guan, S. Yuan, B. Zhang, and Y. D. Chong, Vortex states in an acoustic Weyl crystal with a topological lattice defect, Nat. Commun. **12**, 3654 (2021).

[59] L. Yang, Y. Wang, Y. Meng, Z. Zhu, X. Xi, B. Yan, S. Lin, J. Chen, B. Shi, Y. Ge, S. Yuan, H. Chen, H. Sun, G. Liu, Y. Yang, and Z. Gao, Observation of Dirac Hierarchy in Three-Dimensional Acoustic Topological Insulators, Phys. Rev. Lett. **129**, 125502 (2022).

[60] S. Yin, L. Ye, H. He, X. Huang, M. Ke, W. Deng, J. Lu, and Z. Liu, Valley edge states as bound states in the continuum, Sci. Bull. **69**, 1660 (2024).

[61] A. H. Castro Neto, F. Guinea, N. M. R. Peres, K. S. Novoselov, and A. K. Geim, The electronic properties of graphene, 2009.

[62] C. L. Kane, and E. J. Mele, Quantum Spin Hall Effect in Graphene, Phys. Rev. Lett. **95**, 226801 (2005).

[63] B. A. Bernevig, and S. C. Zhang, Quantum Spin Hall Effect, Phys. Rev. Lett. **96**, 106802 (2006).

[64] Y. Ding, Y. Peng, Y. Zhu, X. Fan, J. Yang, B. Liang, X. Zhu, X. Wan, and J. Cheng, Experimental Demonstration of Acoustic Chern Insulators, Phys. Rev. Lett. **122**, 014302 (2019).

[65] M. Arai, F. Blaschke, M. Eto, and N. Sakai, Massless bosons on domain walls: Jackiw-Rebbi-like mechanism for bosonic fields, Phys. Rev. D **100**, 095014 (2019).

[66] X. Zhang, F. Yu, Z. Chen, Z. Tian, Q. Chen, H. Sun, and G. Ma, Non-Abelian braiding on photonic chips, Nat. Photonics **16**, 390 (2022).

[67] J. Mei, J. Wang, X. Zhang, S. Yu, Z. Wang, and M. Lu, Robust and High-Capacity Phononic

Communications through Topological Edge States by Discrete Degree-of-Freedom Multiplexing, Phys. Rev. Appl. **12**, 054041 (2019).

[68] H. Zhao, X. Qiao, T. Wu, B. Midya, S. Longhi, and L. Feng, Non-Hermitian topological light steering, Science **365**, 1163 (2019).